\documentclass[aps,prb,twocolumn,superscriptaddress,longbibliography,reprint]{revtex4-1}
\usepackage{amsfonts}
\usepackage{amssymb}
\usepackage{amsmath}
\usepackage{graphicx}
\usepackage{epsfig}
\usepackage{subfigure}
\usepackage{appendix}
\usepackage{hyperref}
\usepackage{url}

\hypersetup{hypertex=true,
colorlinks=true,
linkcolor=blue,
urlcolor = blue,
citecolor=blue}

\begin{document}

\title{Magnetization in a non-equilibrium quantum spin system}
\author{X. Z. Zhang}
\email{zhangxz@tjnu.edu.cn}
\affiliation{College of Physics and Materials Science, Tianjin Normal University, Tianjin
300387, China}

\begin{abstract}
The dynamics described by the non-Hermitian Hamiltonian typically capture
the short-term behavior of open quantum systems before quantum jumps occur.
In contrast, the long-term dynamics, characterized by the Lindblad master
equation (LME), drive the system towards a non-equilibrium steady state
(NESS), which is an eigenstate with zero energy of the Liouvillian
superoperator, denoted as $\mathcal{L}$. Conventionally, these two types of
evolutions exhibit distinct dynamical behaviors. However, in this study, we
challenge this common belief and demonstrate that the effective
non-Hermitian Hamiltonian can accurately represent the long-term dynamics of
a critical two-level open quantum system. The criticality of the system
arises from the exceptional point (EP) of the effective non-Hermitian
Hamiltonian. Additionally, the NESS is identical to the coalescent state of
the effective non-Hermitian Hamiltonian. We apply this finding to a series
of critical open quantum systems and show that a local dissipation channel
can induce collective alignment of all spins in the same direction. This
direction can be well controlled by modulating the quantum jump operator.
The corresponding NESS is a product state and maintains long-time coherence,
facilitating quantum control in open many-body systems. This discovery paves
the way for a better understanding of the long-term dynamics of critical
open quantum systems.
\end{abstract}

\maketitle

\section{Introduction}

\label{int} Open quantum many-body systems have emerged as a captivating
research field at the intersection of theoretical and experimental physics
\cite{Breuer2002,Weiss2012,Rivas2012}. Comprising numerous interacting
quantum particles, these systems exhibit intricate and captivating dynamics
that elude traditional closed quantum systems. The interaction of these
systems with an external environment leads to dissipation and decoherence,
presenting new challenges and opportunities for exploring quantum phenomena
\cite{Pellizzari1995,Balasubramanian2009,Lanyon2011,Paik2011}. Recent
advancements have been made in the realization and manipulation of open
quantum many-body systems in atomic, molecular, and optical (AMO) systems
\cite%
{Kasprzak2006,Bloch2008,Bloch2008a,Diehl2008,Syassen2008,Baumann2010,Barreiro2011,Schauss2012,Ritsch2013,Carusotto2013,Daley2014}%
, which offer precise control over individual quantum particles and enable
the engineering of complex interactions and dissipation mechanisms. In
addition, state-of-the-art measurement techniques, such as quantum state
tomography and quantum non-demolition measurements, provide unprecedented
opportunities to investigate the dynamics of these systems with high
precision \cite%
{Nelson2007,Gericke2008,Hofferberth2008,Bakr2009,Sherson2010,Miranda2015,Cheuk2015,Parsons2015,Haller2015,Omran2015,Edge2015,Yamamoto2016,Alberti2016}%
.

The dynamics of an open quantum system are typically described by a quantum
master equation, specifically the Lindblad master equation (LME). This is
attributed to the weak coupling and separation of timescales between the
system and its environment. The Liouvillian superoperator $\mathcal{L}$
governs the time evolution of the density matrix, fully characterizing the
relaxation dynamics of an open quantum system through its complex spectrum
and eigenmodes \cite{Rivas2012}. A notable feature of open quantum systems
is the presence of long-lived states that emerge far from equilibrium, known
as non-equilibrium steady states (NESS). These NESS can exhibit novel
properties, such as the presence of quantum correlations and the breakdown
of conventional statistical mechanics \cite{Eisert2015}. Investigating the
conditions and properties of NESS is currently an active area of research.
\begin{figure}[tbh]
\centering\includegraphics[width=0.48\textwidth]{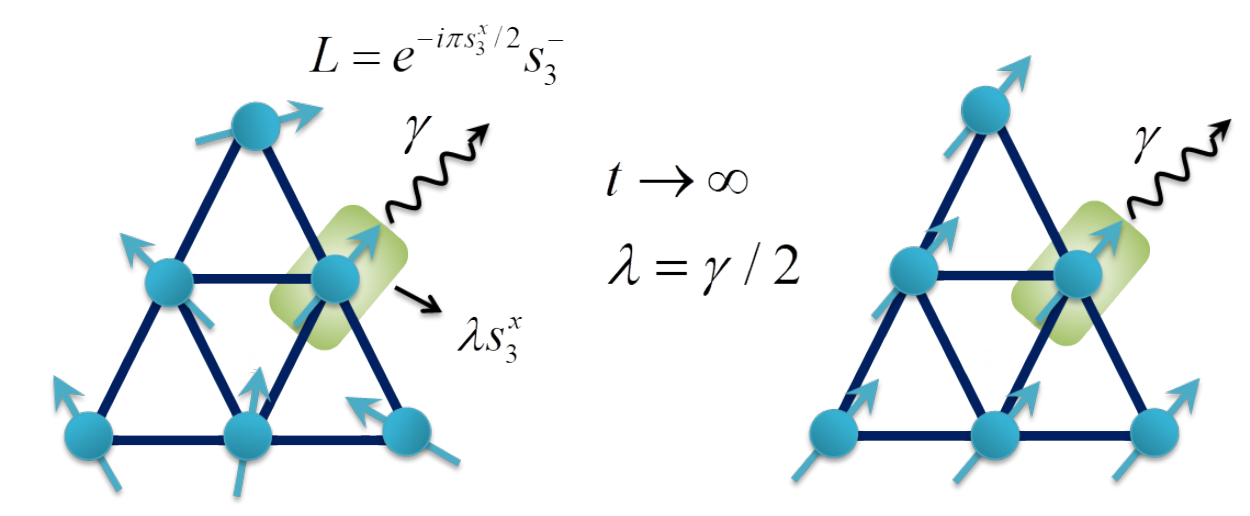}
\caption{Schematic illustration of the magnetization in a non-equilibrium
quantum spin system. The system comprises six spins, with the third spin
subjected to a local external field represented as $\protect\lambda s_{3}^{x} $, and a dissipation channel characterized by the quantum jump
operator $L=e^{-i\protect\pi s^{x}/2}s_{3}^{-}$. The green-shaded region in
the illustration depicts the local external field. The long-lived NESS of
the system converges to a coherent product state, with all spins aligning in
the same direction, when a critical field is activated. This behavior
results from the identical dynamic effects induced by two types of
probability of the the SSE in each quantum trajectory. Importantly, this
finding remains robust irrespective of the system's structure and the
initial spin configuration.} \label{fig_illu}
\end{figure}
The non-Hermitian Hamiltonian is an extension of standard quantum mechanics
that allows for the description of dissipative systems in a minimalistic
manner. In recent years, there has been a growing interest in using
non-Hermitian descriptions to study condensed matter systems \cite%
{Lee2016,Kunst2018,Yao2018,Gong2018,El-Ganainy2018,Nakagawa2018,Shen2018,Wu2019,Yamamoto2019,Song2019,Yang2019,Hamazaki2019,Li2019,Kawabata2019,Kawabata2019a,Lee2019,Yokomizo2019,Jin2020}
. These descriptions have not only expanded the realm of condensed-matter
physics, providing insightful perspectives, but also offered a fruitful
framework for understanding inelastic collisions \cite{Xu2017}, disorder
effects \cite{Shen2018,Hamazaki2019}, and system-environment couplings \cite%
{Nakagawa2018,Yang2019,Song2019}. The interplay between non-Hermiticity and
interactions can lead to exotic quantum many-body effects, such as
non-Hermitian extensions of the Kondo effect \cite%
{Nakagawa2018,Lourenfmmodeboxclsecio2018}, many-body localization \cite%
{Hamazaki2019}, and fermionic superfluidity \cite{Yamamoto2019,Okuma2019}.
One intriguing feature of non-Hermitian systems is the presence of
exceptional points (EPs), which are degeneracies of non-Hermitian operators
where the eigenvalues and corresponding eigenstates merge into a single
state \cite{Berry2004,Heiss2012,Lee2016,Miri2019,Zhang2020}. These EPs give
rise to fascinating dynamical phenomena, including asymmetric mode switching
\cite{Doppler2016}, topological energy transfer \cite{Xu2016}, robust
wireless power transfer \cite{Assawaworrarit2017}, and enhanced sensitivity
\cite{Wiersig2014,Wiersig2016,Hodaei2017,Chen2017}, depending on the nature
of their EP degeneracies. High-order EPs, where more than two eigenstates
coalesce, have attracted significant attention due to their topological and
distinct dynamical properties \cite%
{Zhang2012,Zhang2020a,Zhang2020b,Zhang2020c,Zhang2021,Yang2021,
Xu2023,Xu2023a}.

In the context of open quantum systems, the evolved density matrix driven by
the LME can be obtained by averaging an ensemble of quantum trajectories.
Each trajectory is determined by the stochastic Schr\"{o}inger equation
(SSE). The SSE involves two types of probability evolution: a non-unitary
evolution determined by the effective non-Hermitian Hamiltonian and a state
collapse induced by the quantum jump operator. Generally, the non-Hermitian
Hamiltonian captures the short-term dynamics of the open quantum system
before a quantum jump occurs or describes a post-selected trajectory that
necessitates substantial experimental resources. The dynamical consequences
of the effective non-Hermitian system are irrelevant to the NESS of the open
quantum system. The objective of this paper is to establish the connection
between the dynamics determined by the non-Hermitian Hamiltonian and the
LME. First, we review the connection between the stochastic SSE and the LME.
We demonstrate how to modulate the quantum jump operator to make the evolved
state converge to the coalescent state determined by the critical
non-Hermitian Hamiltonian in each quantum trajectory. Essentially, the
evolution direction dictated by the critical non-Hermitian Hamiltonian
coincides with that determined by the quantum jump operator. This is a
unique characteristic of the critical non-Hermitian Hamiltonian that lacks
in non-Hermitian Hamiltonians without exceptional points (EPs) or imaginary
energy levels. Furthermore, we generalize this mechanism to the critical
quantum spin system with high-order EPs. We demonstrate that a single local
dissipation can cause the collective rotation of spins in a specific
direction, which is shown in Fig. \ref{fig_illu}. The achieved NESS is
equivalent to the coalescent state of such a critical non-Hermitian
Hamiltonian. Remarkably, this non-equilibrium behavior remains unaffected by
the system's geometry, initial spin configuration, and weak disorder, thus
highlighting its robustness. These analytical findings possess independent
interest and hold the potential to inspire future analytical studies on
critical open quantum systems.

The remainder of the paper is organized as follows: Sec. ~\ref{HD} provides
a review of the LME and the SSE, demonstrating the underlying mechanism
using a two-level open quantum system. Sec. \ref{ML} applies the obtained
mechanism to an open quantum spin system. We showcase the coincidence
between EP dynamics and the magnetization of the open quantum spin system.
Furthermore, we analyze the proposed scheme across various system
parameters. We conclude the paper in Sec. \ref{summary}. Supplementary
details of our calculation are provided in the Appendix.

\begin{figure}[tbh]
\centering\includegraphics[width=0.45\textwidth]{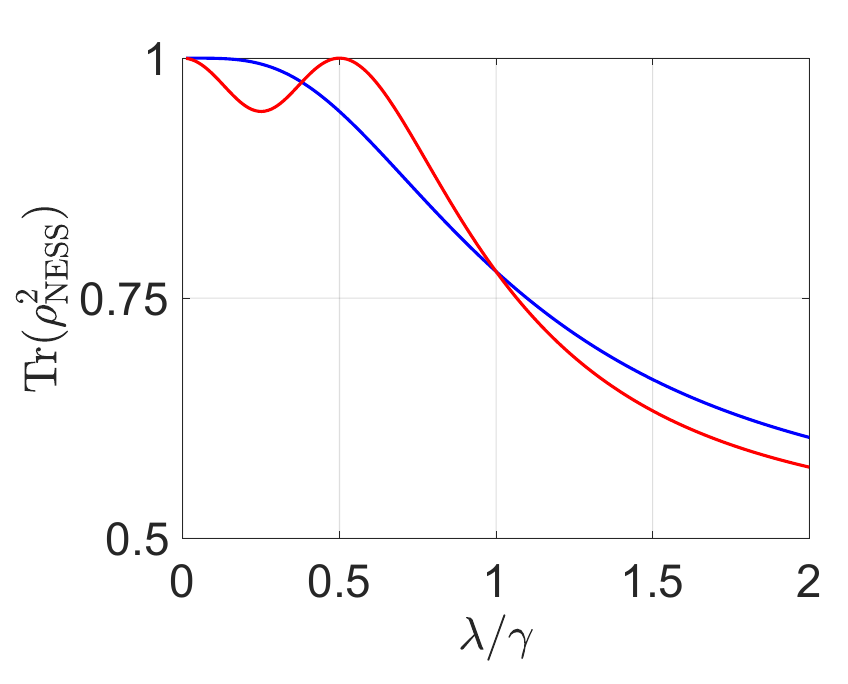}
\caption{Plot of the purity Tr$\left( \protect\rho _{\mathrm{NESS}}^{2}\right) $ as the function of $\protect\lambda /\protect\gamma $. The
blue and red lines correspond to the quantum jump operators $L=s^{-} $, and $\widetilde{L}=e^{-i\frac{\protect\pi }{2}s^{x}}s^{-}$, respectively. The
blue line monotonically decreases to $0.5$, indicating a completely mixed
state. The red line initially decreases and then returns to $1$. There is a
range around $0.5$ where the evolved state can be approximated as a pure
state. When $\protect\lambda /\protect\gamma =0.5$, the NESS is $\protect\rho _{\mathrm{NESS}}$ $|y,+\rangle \langle y,+|$, which is also the
coalescent of $\mathcal{H}$.} \label{fig1}
\end{figure}
\section{Heuristic derivation}

\label{HD} The dynamics of open quantum systems coupled to a Markovian
environment are commonly described by the LME. The equation describing the
time evolution of the density matrix $\rho $ is given by
\begin{equation}
\frac{\text{\textrm{d}}\rho }{\text{\textrm{d}}t}=-i(\mathcal{H}\rho -\rho
\mathcal{H}^{\dagger })+\sum_{\mu }\Gamma _{\mu }L_{\mu }\rho L_{\mu
}^{\dagger }\equiv \mathcal{L}\rho .  \label{LME}
\end{equation}%
In this equation, $\rho $ represents the density matrix. The non-Hermitian
Hamiltonian $\mathcal{H}$ is given by $\mathcal{H}$ $\mathcal{=}$ $H-\frac{i%
}{2}\sum_{\mu }\Gamma _{\mu }L_{\mu }^{\dagger }L_{\mu }$, where $H$ is a
Hermitian operator representing the system Hamiltonian. The non-Hermitian
nature of $\mathcal{H}$ accounts for the non-unitary dynamics observed in
open quantum systems. The jump operators $L_{\mu }$ describe the dissipative
quantum channels with a strength of $\Gamma _{\mu }$. $\mathcal{L}$ is the
Liouvillian superoperator. Alternatively, one can track the trajectory of a
pure state using a SSE, such as
\begin{eqnarray}
\text{\textrm{d}}|\Psi \rangle &=&-i\mathcal{H}|\Psi \rangle \text{\textrm{d}%
}t+\frac{1}{2}\sum_{\mu }\Gamma _{\mu }[\langle \Psi |L_{\mu }^{\dagger
}L_{\mu }|\Psi \rangle ]|\Psi \rangle \text{\textrm{d}}t  \notag \\
&&+\sum_{\mu }(\frac{L_{\mu }|\Psi \rangle }{\sqrt{\langle \Psi |L_{\mu
}^{\dagger }L_{\mu }|\Psi \rangle }}-|\Psi \rangle )\text{\textrm{d}}N_{\mu
},  \label{SSE}
\end{eqnarray}%
where the Poisson increment \textrm{d}$N_{\mu }$ satisfies \textrm{d}$N_{\mu
}$\textrm{d}$N_{\nu }=\delta _{\mu \nu }$, taking the value $0$ or $1$. The
jump operators in the LME correspond to the stochastic jumps in the SSE. If
\textrm{d}$N_{\mu }=0$, the evolution is solely described by the
non-Hermitian Hamiltonian $\mathcal{H}$, which is referred to as the
no-click limit \cite{Daley2014}. However, this limit is rarely achieved in
experiments since its realization requires exponentially many experiments to
be carried out before a desired trajectory is obtained. The
connection between the SSE and the LME lies in the relationship
between the individual wave function trajectories and the ensemble-averaged
density matrix. By averaging over the different realizations of the
stochastic trajectories generated by the SSE, one can recover the
ensemble-averaged dynamics described by the LME. In this way, the SSE
provides a more detailed and microscopic description of the dynamics, while
the LME provides a coarse-grained description that captures the averaged
behavior of the system \cite{Daley2014}.

To fully grasp the essence of this paper, we begin by considering a simple
quantum system comprising two levels with orthonormal states. This model has
diverse applications and can describe phenomena such as the spin degree of
freedom of an electron, a simplified representation of an atom with only two
atomic levels, the lowest eigenstates of a superconducting circuit, or the
discrete charge states of a quantum dot. In this model, the system
Hamiltonian is given by $H=\lambda s^{x}$, where $\lambda $ represents the
energy difference between the two states and $\sigma ^{x}=2s^{x}$ is the
Pauli matrix corresponding to the $x$-direction. The quantum jump operator
is denoted as $L_{\mu }=s^{-}$ with a strength $\Gamma _{\mu }=\gamma $,
where $s^{-}$ represents the lowering operator responsible for the
spin flip from the spin-up state to the spin-down state. The initial state $%
|\Psi \left( 0\right) \rangle $ is assumed to be an arbitrary pure state
applicable to various many-body examples. In this context, the initial state
is represented by the density matrix $\rho \left( t=0\right) =|\Psi \left(
0\right) \rangle \langle \Psi \left( 0\right) |$. Referring to the Eq. (\ref%
{SSE}), the evolution of $|\Psi \left( t+\delta t\right) \rangle $ is
determined by either $\frac{\left( 1-i\mathcal{H}\delta t\right) }{\sqrt{%
1-\delta p}}|\Psi \left( t\right) \rangle $ with a probability of $1-\delta
p $ or $\frac{L}{\sqrt{\delta p/\delta t}}|\Psi \left( t\right) \rangle $
with a probability of $\delta p$. Here $\delta p$ is red defined as
\begin{equation}
\delta p=\langle \Psi \left( t\right) |s^{+}s^{-}|\Psi \left( t\right)
\rangle \delta t.
\end{equation}

Notice that when $\lambda >\gamma /2$, the state $|\Psi \left( t\right)
\rangle $ oscillates between the two eigenstates of the non-Hermitian
Hamiltonian $\mathcal{H}$ , which possesses a full real spectrum except for
a common imaginary part eliminated by the amplitude $1/\sqrt{1-\delta p}$.
On the other hand, if $\lambda <\gamma /2$, $|\Psi \left( t\right) \rangle $
relaxes to the eigenstate with the maximum imaginary part, as $\mathcal{H}$
has two complex eigenvalues. It is worth noting that when $\lambda =
\gamma/2 $, an exceptional point (EP) exists in the spectrum of $\mathcal{H}$%
, where there is only one coalescent eigenstate $|\psi_{\mathrm{c}}\rangle =
\frac{1}{\sqrt{2}}\left(1, i\right)^T$. For an arbitrary initial state $%
|\Psi \left( 0\right) \rangle $, it evolves towards the coalescent state $%
|\psi _{\mathrm{c}}\rangle $ due to the nilpotent matrix property of $%
\mathcal{H}$, i.e., $\mathcal{H}^{2}=0$ (see Appendix \ref{EP_two} for more
details). Alternatively, if the final state is a steady pure state, it can
be projected onto the Bloch sphere, revealing a definite spin direction.
However, the presence of the quantum jump operator $s^{-}$ disrupts the
evolution driven by $\mathcal{H}$ and consequently affects the direction.
The steady state must strike a balance between these two probabilistic
evolutions. To gain further insight into the NESS $\rho _{\mathrm{NESS}}$
defined by \textrm{d}$\left( \rho _{\mathrm{NESS}}\right) /$\textrm{d}$t=0$,
we employ a spin bi-base mapping, also known as the Choi-Jamio\l kowski
isomorphism, to map a density matrix to a vector in the computational bases
(see Appendix \ref{NESS_two} for more details). The NESS corresponds to the
eigenstate of $\mathcal{L}$ with zero eigenvalue, which can be expressed as%
\begin{equation}
\rho _{\mathrm{NESS}}=\left(
\begin{array}{cc}
\frac{\lambda ^{2}}{2\lambda ^{2}+\gamma ^{2}} & -i\frac{\lambda \gamma }{%
2\lambda ^{2}+\gamma ^{2}} \\
i\frac{\lambda \gamma }{2\lambda ^{2}+\gamma ^{2}} & \frac{\lambda
^{2}+\gamma ^{2}}{2\lambda ^{2}+\gamma ^{2}}%
\end{array}%
\right) .
\end{equation}

The coherence of a system can be measured by its purity, which is quantified
by the function Tr$(\rho_\mathrm{NESS}^2)$. In Fig. \ref{fig1}, we depict the
behavior of Tr$(\rho_\mathrm{NESS}^2)$ as a function of $\lambda$, while
keeping $\gamma$ fixed at 1. Let us first consider two limiting cases: When $%
\lambda = 0$, the non-Hermitian Hamiltonian $\mathcal{H}$ drives the initial
state to $|\psi_\mathrm{f}\rangle = (0,1)$. Simultaneously, the quantum jump
operator projects the spin to the down state. Consequently, $|\psi_\mathrm{f}%
\rangle$ becomes the non-equilibrium steady state (NESS). On the other hand,
when $\lambda \gg \gamma$, the density matrix $\rho_\mathrm{NESS}$
simplifies to
\begin{equation}
\rho _{\mathrm{NESS}}=\left(
\begin{array}{cc}
1/2 & 0 \\
0 & 1/2%
\end{array}%
\right) ,
\end{equation}%
which corresponds to a completely mixed state. This can be understood as
follows: Under the influence of the non-Hermitian Hamiltonian $\mathcal{H}$,
the evolved state does not have a definite direction. Instead, it oscillates
between the two eigenvectors along the $x$-direction, i.e., $|\psi _{\mathrm{%
1}}\rangle =\left( 1,\text{ }1\right) ^{T}/\sqrt{2}$ and $|\psi _{\mathrm{2}%
}\rangle =\left( 1,\text{ }-1\right) ^{T}/\sqrt{2}$. However, the quantum
jump operator forces the spin to align parallel to the $-z$-direction. The
consequences of the two effects are completely independent and cannot be
reconciled, leading to a thermal state with infinite temperature. Generally,
the NESS is not a pure state, except for a few limiting cases. Therefore the
evolution of the state cannot be mapped onto the Bloch sphere, making it
impossible to analyze its trajectory on the sphere. However, by applying a
rotation to the quantum jump operator, i.e., $\widetilde{L}=Us^{-}$, where $%
U=e^{-i\frac{\pi }{2}s^{x}}$ corresponds to a unitary feedback operator \cite%
{Lloyd2000,Nelson2000}, the new quantum jump operator $\widetilde{L}$ drives
the evolved state $|\Psi \left( t\right) \rangle $ towards $|y,+\rangle
=|\psi _{\mathrm{c}}\rangle $, which is the eigenstate of the operator $%
s^{y} $ with eigenenergy $1/2$. Importantly, the probability $\delta p$ and
the non-Hermitian Hamiltonian $\mathcal{H}$ remain unchanged since $%
\widetilde{L}^{\dagger }\widetilde{L}=s^{+}s^{-}=L^{\dagger }L$. When $%
\lambda =\gamma /2$, $|y,+\rangle $ also represents the coalescent state of $%
\mathcal{H}$. The EP dynamics guides the evolved state $|\Psi \left(
t\right) \rangle $ towards $|y,+\rangle $. These two probabilistic
evolutions tend to drive the arbitrary initial state $|\Psi \left( 0\right)
\rangle $ to $|y,+\rangle $, resulting in the final steady state being a
pure state $|y,+\rangle $. This can be demonstrated in Fig. \ref{fig1}. The
purity of $\rho _{\mathrm{NESS}}$, represented by the red line, initially
decays and then recovers to $1$ when $\lambda /\gamma =1/2$. At this point, $%
\rho _{\mathrm{NESS}}=\left( I_{2}+\sigma ^{y}\right) /2=|y,+\rangle \langle
y,+|$, which validates our previous analysis. Based on the above
calculations, one can appropriately choose the quantum jump operator to
achieve the desired spin polarization.

\section{Magnetization in open quantum spin systems induced by a local
dissipation channel}

\label{ML}
\begin{figure}[tbh]
\centering\includegraphics[width=0.45\textwidth]{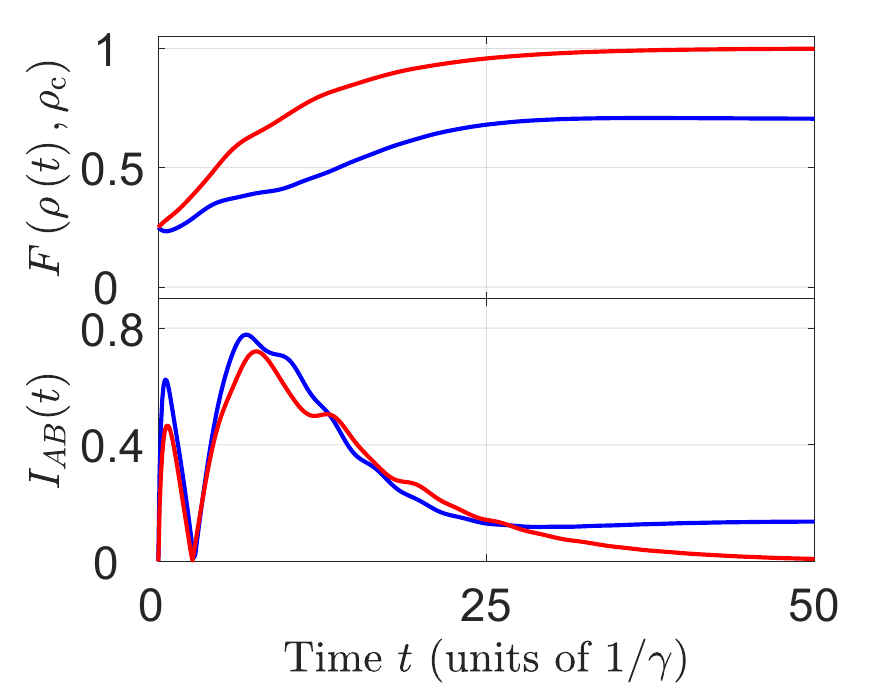}
\caption{The time evolutions of the Uhlmann fidelity $F(\protect\rho(t),
\protect\rho_\mathrm{c})$ and quantum mutual information $I_{AB}(t) $. The
blue and red lines represent the driven systems under the quantum jump
operators $L_{1}$ and $\widetilde{L}_{1}$, respectively. The system
parameters are chosen as $\protect\lambda/\protect\gamma = 0.5$ and $J_{ij}/\protect\gamma = 2$. The system is initially prepared in the state $|\Downarrow\rangle$ such that $F(\protect\rho(0), \protect\rho_\mathrm{c}) =
1/N$, where $N = 4$ for simplicity. In the absence of modulation, the
Uhlmann fidelity $F(\protect\rho(t), \protect\rho_\mathrm{c})$ (blue curve)
saturates at approximately 0.75. On the other hand, the local dissipation
channel $\widetilde{L}_{1}$ drives the system towards the state $\protect\rho_\mathrm{c}$ with zero quantum mutual information $I_{AB}(t\rightarrow
\infty) = 0$. It is noteworthy that these conclusions hold irrespective of
the system's configuration and size.} \label{fig2}
\end{figure}

In this section, we extend our main conclusion to a many-body quantum spin
system based on the mechanism described above. In this case, the system is
assumed to be described by a Heisenberg Hamiltonian under the influence of
an external field. The Hamiltonian $H$ is defined as follows:
\begin{eqnarray}
H &=&H_{\text{\textrm{spin}}}+H_{\mathrm{e}}, \\
H_{\text{\textrm{spin}}} &=&-\sum_{i,j\neq i}(J_{ij}/2)\left(
s_{i}^{+}s_{j}^{-}+s_{i}^{-}s_{j}^{+}\right) +\sum_{i,j\neq i}\Delta
_{ij}s_{i}^{z}s_{j}^{z},  \notag \\
H_{\mathrm{e}} &=&\sum_{i}\lambda _{i}\mathbf{h}\cdot \mathbf{s}_{i}.
\end{eqnarray}%
Here the operators $s_{i}^{\pm }=s_{i}^{x}\pm is_{i}^{y}$ and $s_{i}^{z}$
represent spin-$1/2$ operators at the $i$-th site, which obey the standard
SU(2) symmetry relations: $[s_{i}^{z},s_{j}^{\pm }]=\pm s_{i}^{\pm }\delta
_{ij}$ and $[s_{i}^{+},s_{j}^{-}]=2s_{i}^{z}\delta _{ij}$, where $\delta
_{ij}$ is the Dirac delta function. The summation $\sum_{i,j\neq i}$ implies
the summation over of possible pair interactions within an arbitrary range.
The parameter $J_{ij}$ represents the inhomogeneous spin-spin interaction,
while $\Delta _{ij}$ characterizes the anisotropy of the spin system $H_{%
\text{\textrm{spin}}}$. The local external field $\mathbf{h} = (1, 0, 0)$
can be interpreted as a magnetic field along the $x$-direction and is
experimentally accessible in ultracold atom experiments \cite%
{Lee2014,Ashida2017,Pan2019b}. The strength experienced by each spin is
denoted as $\lambda _{i}$. When $\Delta _{ij}=J_{ij}/2$, the system $H_{%
\text{\textrm{spin}}}$ corresponds to a ferromagnetic Heisenberg Hamiltonian
that respects the SU(2) symmetry, i.e., $[\sum_{i}s_{i}^{\sigma },H_{\text{%
\textrm{spin}}}]=0$ with $\sigma =\pm ,z$. Thus, the eigenstates of $H_{%
\text{\textrm{spin}}}$ can be classified based on the total spin number $s$.
Among these states, a fully polarized ferromagnetic state, denoted as $%
\left\vert \Uparrow \right\rangle =\prod\limits_{i=1}^{N}\left\vert \uparrow
\right\rangle _{i}$ belongs to the ground states multiplet, where $|\uparrow
\rangle _{i}\left( |\downarrow \rangle _{i}\right) $ is the eigenstate of $%
s_{i}^{z}$ with eigenenergy $\frac{1}{2}(-\frac{1}{2})$ \cite%
{Heisenberg1928,Yang1966}. The degenerate ground states $\left\{ \left\vert
G_{n}\right\rangle \right\} $ belonging to the subspace $s=N/2$ are given by
$\left\vert G_{n}\right\rangle =(\sum_{i}s_{i}^{-})^{n-1}\left\vert \Uparrow
\right\rangle $, where $n$ ranges from $1$ to $N+1$. Clearly, $\left\{
\left\vert G_{n}\right\rangle \right\} $ are the degenerate groundstates of $%
H_{\text{\textrm{spin}}}$ with an $(N+1)$-fold degeneracy, where all the
spins are aligned in the same direction. However, the presence of the
external field $H_{\mathrm{e}}$ breaks the SU(2) symmetry of the system,
consequently splitting the degeneracy of these states. From this point
onward, we will assume $\Delta_{ij} = J_{ij}/2$ for clarity.
\begin{figure}[tbh]
\centering\includegraphics[width=0.45\textwidth]{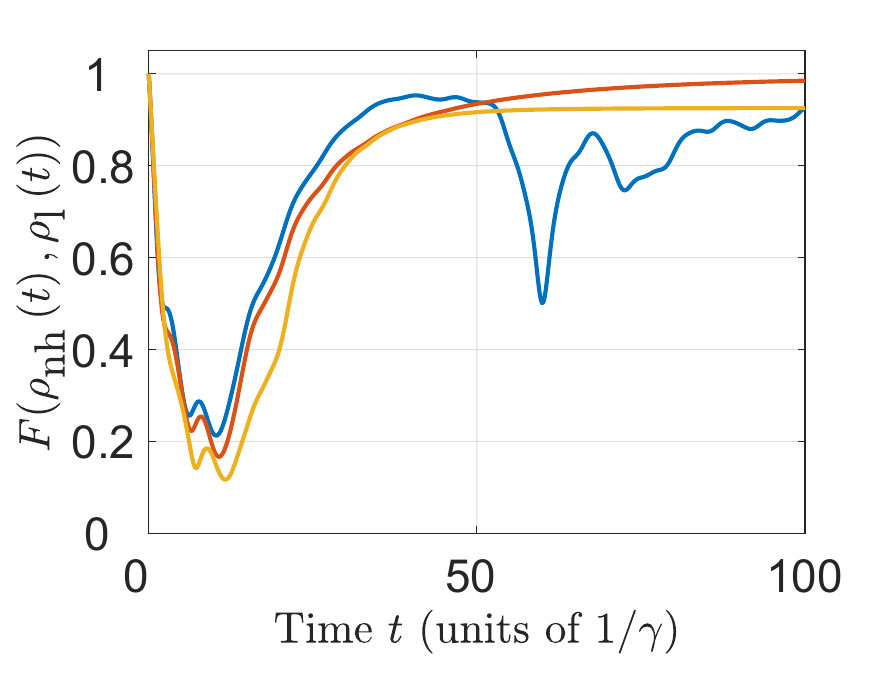}
\caption{Plots of Uhlmann fidelity $F(\protect\rho _{\text{\textrm{nh}}}\left( t\right) ,\protect\rho _{\text{\textrm{l}}}\left( t\right) )$ as a
function of time $t$. Both $\protect\rho _{\text{\textrm{nh}}}\left(
t\right) $ and $\protect\rho _{\text{\textrm{l}}}\left( t\right) $ are
initialized in the state $\left\vert \Downarrow \right\rangle $ and driven
by the effective non-Hermitian Hamiltonian $\mathcal{H}_{\mathrm{spin}}$ and
$\mathcal{L}$. The blue, red, and yellow lines correspond to values of $\protect\lambda/\protect\gamma$ equal to $2/3$, $1/2$, and $1/4$,
respectively. At $\protect\lambda/\protect\gamma = 1/2$, which corresponds
to the EP of $\mathcal{H}{\text{spin}}$, $F(\protect\rho _{\text{\textrm{nh}}}\left( t\right) ,\protect\rho _{\text{\textrm{l}}}\left( t\right) )$
approaches $1$ since the coalescent state $\protect\rho _{\mathrm{c}}$ is
the final steady state for $\mathcal{L}$. For $\protect\lambda /\protect\gamma >1/2$, a full real spectrum of $\mathcal{H}_{\mathrm{spin}}$ emerges
leading to the periodic evolution of $\protect\rho _{\text{\textrm{nh}}}\left( t\right) $ without a definite direction in the Bloch sphere. This
oscillatory behavior is represented by the blue line. When $\protect\lambda/\protect\gamma < 1/2$, $\mathcal{H}{\text{spin}}$ exhibits imaginary energy
levels, and the final steady state is determined by the maximum value among
them. However, this state does not coincide with $\protect\rho_{\text{l}}(t\rightarrow \infty)$ due to the different effects of the jump operator $L_{1}$ and $\mathcal{H}{\text{spin}}$. Consequently, $F(\protect\rho_{\text{nh}}(t\rightarrow \infty), \protect\rho_{\text{l}}(t\rightarrow \infty))<1$,
indicating a deviation from unity. This can be observed in the plot
represented by the yellow line.} \label{fig3}
\end{figure}

The dissipation channels $L_{i}=s_{i}^{-}$ are now applied to all the local
sites under the influence of a magnetic field. For simplicity, we will focus
on the case where only a single lattice site is affected by the external
field and dissipation channel. Specifically, we set $\lambda _{i}=\lambda
\delta _{i,1}$ and $\Gamma _{\mu }=\gamma \delta _{\mu ,1}$. The extension
to multiple lattice sites is straightforward. Following Eq. (\ref{SSE}), we
can divide the dynamics into two parts: the first part involves the quantum
jump operator that flips the spin on the first site from the up state $%
|\uparrow \rangle _{1}$ to the down state $|\downarrow \rangle _{1}$.
Considering the external field as a perturbation, the low-energy excitation
of the ferromagnetic Heisenberg model can be described by magnons.
Intuitively, the collective behavior of spins leads to the spreading of the
effect of $s_{1}^{-}$ across the entire system, ultimately resulting in the
attainment of the final steady state $\left\vert \Downarrow \right\rangle $.
The second part characterizes the non-unitary dynamics driven by the
non-Hermitian spin Hamiltonian
\begin{equation}
\mathcal{H}_{\mathrm{spin}}=H-i\gamma s_{1}^{+}s_{1}^{-}/2.  \label{non_spin}
\end{equation}%
Clearly, the local external field $H_{\mathrm{e}}=\lambda s_{1}^{x}$ and
on-site dissipation channel $-i\gamma s_{1}^{+}s_{1}^{-}/2$ can be combined
into a complex filed $H_{\mathrm{ec}}=\lambda s_{1}^{x}-i\gamma
s_{1}^{+}s_{1}^{-}/2$ applied to the ferromagnetic Heisenberg Hamiltonian $%
H_{\text{\textrm{spin}}}$. In general, the commutation relation $[H_{\text{%
\textrm{spin}}},$ $H_{\mathrm{ec}}]\neq 0$ leads to a splitting of the
ground state of $H_{\text{\textrm{spin}}}$ under the influence of $H_{%
\mathrm{ec}}$. However, when $\lambda \rightarrow \gamma /2$, the spliting
approaches $0$, allowing us to treat $H_{\mathrm{ec}}$ as a non-Hermitian
perturbation. To facilitate this treatment, we introduce the unitary
transformation $U=\prod\nolimits_{j}U^{j}$ with $U^{j}=e^{-i\pi s_{j}^{x}/2}$%
, which represents a collective spin rotation along the $s^{x}$ direction by
an angle $\pi /2$. The matrix form of $H_{\mathrm{ec}}$ in the degenerate
subspace spanned by $\{|\widetilde{G}_{n}\rangle \}=\{U|G_{n}\rangle \}$ can
be given as
\begin{eqnarray}
W_{m,n} &=&\sqrt{\left( N+1-m\right) m}[\left( \lambda -\gamma /2\right)
\delta _{m+1,n}  \notag \\
&&+\left( \lambda +\gamma /2\right) \delta _{m,n+1}]/2N.
\end{eqnarray}%
Here, $W_{m,n}=\langle \widetilde{G}_{m}|UH_{\mathrm{ec}}U^{-1}|\widetilde{G}%
_{n}\rangle $. When $\lambda =\gamma /2$, it reduces to a Jordan block
matrix with an EP of ($N+1$) order. The corresponding coalescent state with
geometric multiplicity of $1$ is given as
\begin{equation}
|\widetilde{G}_{1}\rangle =\prod\nolimits_{j}|y,+\rangle _{j},
\end{equation}%
which represents all the spins aligning parallel to the $+y$ direction.
These results are detailed and exemplified in the Appendix \ref{NHH}. For an
arbitrary initial state $\sum_{n}c_{n}\left( 0\right) |\widetilde{G}%
_{n}\rangle $ within the subspace $s=N/2$, the coefficient $c_{m}\left(
t\right) $ is given by the EP dynamics as
\begin{eqnarray}
c_{m}\left( t\right) &=&c_{m}\left( 0\right) +\sum_{n\neq m}\left( \frac{%
-it\lambda }{N}\right) ^{m-n}\frac{h\left( m-n\right) }{\left( m-n\right) !}
\notag \\
&&\times \lbrack \prod\limits_{p=n+1}^{m}p\left( N+1-p\right)
]^{1/2}c_{n}\left( 0\right) ,
\end{eqnarray}%
where $h\left( m-n\right) $ is the Heaviside step function (refer to
Appendix \ref{NHH} for more details). The expression shows that the
coefficient $c_{N+1}\left( t\right) $ of the evolved state always contains
the highest power of time $t$. As a result, the component $c_{N+1}\left(
t\right) $ dominates over the other components, leading to the final steady
state being the coalescent state $|\psi _{\mathrm{c}}\rangle =e^{-i\frac{\pi
}{2}s^{x}}\left\vert \Downarrow \right\rangle $ with $s^{x}=%
\sum_{i}s_{i}^{x} $. This implies that all spins align in parallel to the $y$%
-direction. We would like to emphasize that while our
primary focus lies on the subspace indexed by $s=N/2$, the critical complex
magnetic field resulting from local dissipation can also lead to the
coalescence of eigenstates in each degenerate subspace with different
quantum number $s$. Consequently, the coalescent states in each subspace
have a geometric multiplicity of $1$. By following the EP$(N+1)$ dynamics in
the $s=N/2$ subspace, an arbitrary initial state in a $s\neq N/2$ subspace
evolves towards the corresponding coalescent state. If the initial state
consists of multiple different types of coalescent states, then the final
state is determined by the coalescent state whose time-dependent coefficient has
the highest power of $t$.
\begin{figure*}[tbh]
\centering\includegraphics[width=0.95\textwidth]{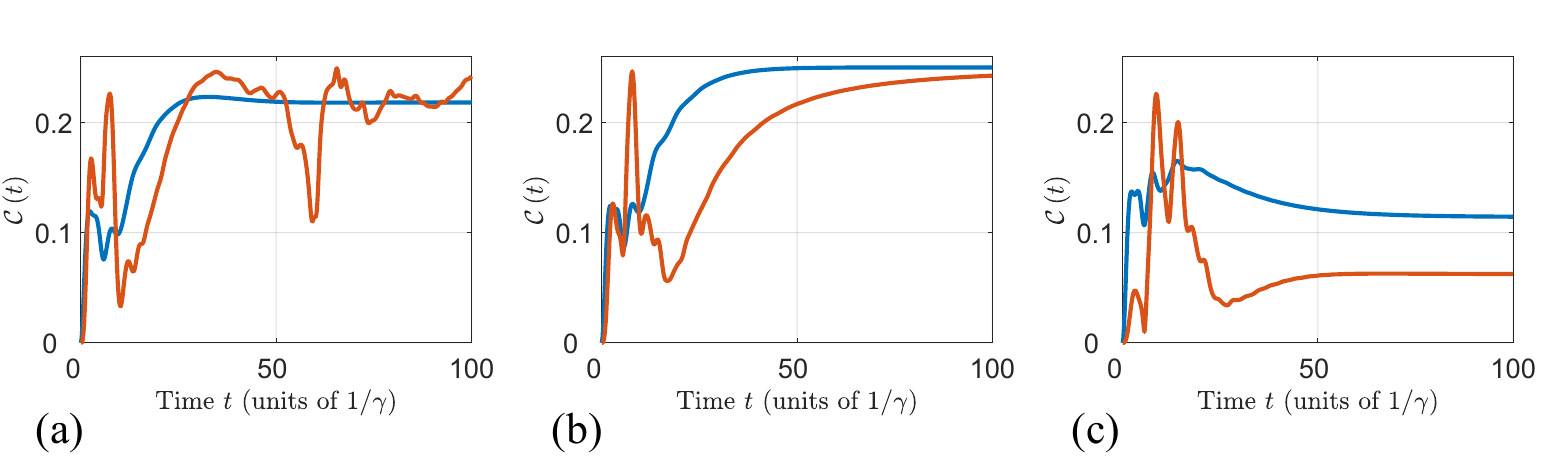}
\caption{Time evolutions of the correlator $\mathcal{C}\left( t\right) $ for
the evolved states $\protect\rho _{\text{\textrm{l}}}\left( t\right) $ and $%
\protect\rho _{\text{\textrm{nh}}}\left( t\right) $. The system parameters
are set as follows: (a) $\protect\lambda /\protect\gamma =2/3$, (b) $\protect%
\lambda /\protect\gamma =1/2$, and (c) $\protect\lambda /\protect\gamma =1/4$%
. In each panel, the blue and red lines represent the $\mathcal{C}(t)$
driven by the Liouvillian superoperator $\mathcal{L}$ and the effective
non-Hermitian Hamiltonian $\mathcal{H}_{\text{spin}}$, respectively. The
correlator $\mathcal{C}\left( t\right) $ driven by $\mathcal{L}$
characterizes the spin-spin correlation of the NESS and asymptotically
approaches a steady value over time. However, in the presence of an
EP or imaginary energy levels in the Hamiltonian $\mathcal{H}_{\mathrm{spin}%
} $, the correlator $\mathcal{C}\left( t\right) $ converges to a specific
asymptotic value instead of exhibiting indefinite oscillation. At the EP ($%
\protect\lambda /\protect\gamma =1/2$), both lines converge to the same
value, $\mathcal{C}(t\rightarrow \infty) = 1/4$, as shown in Fig. 4(b).}
\label{fig4}
\end{figure*}

It can be imagined that the system will not approach the coalescent state $%
|\psi _{\mathrm{c}}\rangle $ under the effect of the Liouvillian
superoperator $\mathcal{L}$ due to the distinct operations of two types of
evolutions. To confirm this conjecture, we introduce Uhlmann fidelity \cite%
{Jozsa1994} which measures the distance between density operators, defined by%
\begin{equation}
F\left( \rho \left( t\right) ,\rho _{\mathrm{c}}\right) =(\text{\textrm{Tr}}%
\sqrt{\sqrt{\rho \left( t\right) }\rho _{\mathrm{c}}\sqrt{\rho \left(
t\right) }})^{2}\text{,}
\end{equation}%
where $\rho _{\mathrm{c}}=|\psi _{\mathrm{c}}\rangle \langle \psi _{\mathrm{c%
}}|$ and $\rho \left( t\right) $ denotes the evolved density matrix. The
system is initialized in the state $\left\vert \Downarrow \right\rangle $.
The second physical quantity of interest is the quantum mutual
information of the bipartite state $\rho _{AB}\left( t\right) $,
which is defined as%
\begin{equation}
I_{AB}=S\left( \rho _{A}\right) +S\left( \rho _{B}\right) -S\left( \rho
_{AB}\right) ,
\end{equation}%
where $S\left( \rho _{\sigma }\right) =-$\textrm{Tr}$\rho _{\sigma }$\textrm{%
ln}$\rho _{\sigma }$ ($\sigma =A,B$) represents the Von Neumann entropy of $%
\rho _{\sigma }\left( t\right) $. This entropy is obtained by tracing out
system $B$ or $A$ from the joint density matrix $\rho _{AB}$. More
specifically, $\rho _{A}\left( t\right) =$\textrm{Tr}$_{B}[\rho _{AB}\left(
t\right) ]$ or $\rho _{B}\left( t\right) =$\textrm{Tr}$_{A}[\rho _{AB}\left(
t\right) ]$. $S\left( \rho _{AB}\right) $ denotes the Von Neumann entropy of
the total state. The quantity $I_{AB}$ is formally equivalent to the
classical mutual information, with the Shannon entropy replaced by its
quantum counterpart. Utilizing $I_{AB}$, we can effectively capture the
separability of the evolved state. If $I_{AB}=0$, the evolved state $\rho
_{AB}\left( t\right) $ is considered simply separable or a product state. In
our system, we divide it into two parts: part $A$ consists
of a single local spin, while part $B$ represents its complement. When the
NESS assumes a product form, $I_{AB}$ will be $0$. In Fig. \ref{fig2}, we
conduct a numerical simulation on these two quantities.  The
results indicate that $F\left( \rho \left( t\right) ,\rho _{\mathrm{c}%
}\right) $ initially increases during the short-time evolution, as it is
determined by the non-Hermitian Hamiltonian $\mathcal{H}_{\mathrm{spin}}$,
which drives $\rho \left( t\right) $ towards $\rho _{\mathrm{c}}$. However, as the long-time evolution progresses, a compromise
between two distinct types of probabilistic evolution emerges, leading to a
deviation of the NESS from $\rho _{\mathrm{c}}$. Additionally, we observe
that the minimum value of $I_{AB}$ is approximately $0.136$, implying that
the spin at the first site remains correlated with the other component.
Consequently, the evolved state $\rho (t)$ is not a product state.

To recover the final state $|\psi _{\mathrm{c}}\rangle $, one should
introduce a unitary operator $U=e^{-i\frac{\pi }{2}s^{x}}$ to the quantum
jump operator
\begin{equation}
\widetilde{L}_{1}=Us_{1}^{-}=\left( s_{1}^{x}-is_{1}^{z}\right) U.
\end{equation}%
The operator $\widetilde{L}$ performs two operations: firstly, it rotates
the spin by an angle of $\pi /2$ along the $x$-direction, and secondly, it
projects the spin at the first site onto the $y$-direction, resulting in the
state $|y_{1},+\rangle $. The unitary operation $U=e^{-i\frac{\pi }{2}s^{x}}$
does not affect non-Hermitian Hamiltonian $\mathcal{H}_{\mathrm{spin}}$
since $\widetilde{L}_{1}^{\dagger }\widetilde{L}_{1}=L_{1}^{\dagger }L_{1}$.
Its effect is limited to the quantum trajectories that deviate from the
post-selected no-click trajectory. However, the effect of $\widetilde{L}_{1}$ on
the first spin is equivalent to that of $\mathcal{H}_{\mathrm{spin}}$ which
tends to freeze each spin along $y$-direction. As a result, regardless of
the type of probabilistic evolution in each quantum trajectory, the
long-term tendency leads to the same consequence, suggesting that $\rho _{%
\mathrm{NESS}}=|\psi _{\mathrm{c}}\rangle \langle \psi _{\mathrm{c}}|$
represents the NESS of the open quantum spin system. In the Appendix \ref%
{NESS_local}, we verify that $\rho _{\mathrm{NESS}}$ is indeed the
eigenfunction of the Liouvillian superoperator $\mathcal{L}$ with zero
energy. Consequently, $\mathcal{H}_{\mathrm{spin}}$ and $\mathcal{L}$ share
the same steady state within the subspace $\left\{ \left\vert
G_{n}\right\rangle \right\} $. This is further confirmed in Fig. \ref{fig2},
where the Uhlmann fidelity $F\left( \rho \left( t\right) ,\rho _{\mathrm{c}%
}\right) \rightarrow 1$ and $I_{AB}\left(t\right) \rightarrow 0$
correspond to the final product state of $|\psi _{\mathrm{c}}\rangle $.
Furthermore, we compare the evolution of two density matrices driven by $%
\mathcal{H}_{\mathrm{spin}}$ and $\mathcal{L}$, at $\lambda =\gamma /2$,
respectively. The initial state is $\left\vert \Downarrow \right\rangle $,
prepared within the subspace $\left\{ \left\vert G_{n}\right\rangle \right\}
$. We examine the Uhlmann fidelity between the two evolved states $\rho _{%
\text{\textrm{nh}}}\left( t\right) $ and $\rho _{\text{\textrm{l}}}\left(
t\right) $ as depicted in Fig. \ref{fig3}, where $\rho _{\text{\textrm{nh}}%
}\left( t\right) =e^{-i\mathcal{H}_{\mathrm{spin}}t}\rho _{\text{\textrm{nh}}%
}\left( 0\right) e^{i\mathcal{H}_{\mathrm{spin}}^{\dagger }t}$ and $\rho _{%
\text{\textrm{l}}}\left( t\right) =e^{\mathcal{L}t}\rho _{\text{\textrm{l}}%
}\left( 0\right) $. The two states $\protect\rho _{\text{\textrm{nh}}}\left(
0\right) $ and $\protect\rho _{\text{\textrm{l}}}\left( 0\right) $ are
initialized in the state $\left\vert \Downarrow \right\rangle $. The fidelity initially decreases
and then rapidly increases to $1$, indicating that the long-time dynamics of
$\rho \left( t\right) $ driven by $\mathcal{L}$ can be effectively described
by $\mathcal{H}_{\mathrm{spin}}$. To gain further insight into the two types of the evolution, we also investigate the time evolution of the correlator $%
\mathcal{C}\left( t\right) =\mathrm{Tr}[\rho \left( t\right)
s_{1}^{+}s_{N}^{-}]$ for two such evolved states. In Fig. \ref{fig4}(b), the two curves exhibit the same long-time tendency and finally approaches $\mathcal{C}\left( t\right) =0.25$
when $\lambda /\gamma =1/2$, which can be also captured by the Uhlmann fidelity. This result is quite astonishing as it challenges the common
belief that $\mathcal{H}_{\mathrm{spin}}$ captures the short-time dynamics
before a quantum jump occurs, while $\mathcal{L}$ characterizes the
long-time dynamics. Before ending this discussion, it is worth
noting that when the initial state is prepared in a different degenerate
subspace ($s\neq N/2$), the final evolved state becomes an entangled state
rather than a separable state where all the spins align in the same
direction, as observed in the $s=N/2$ subspace. Achieving collective
magnetization requires careful modulation of the quantum jump operator $%
\mathcal{L}$ to align its action with the effect of $\mathcal{H}_{\mathrm{%
spin}}$. This process may involve multiple dissipation channels and present
significant challenges in both theoretical and experimental aspects.
Consequently, our proposal is specifically applicable to the $s=N/2$
subspace.
\begin{figure}[tbh]
\centering\includegraphics[width=0.45\textwidth]{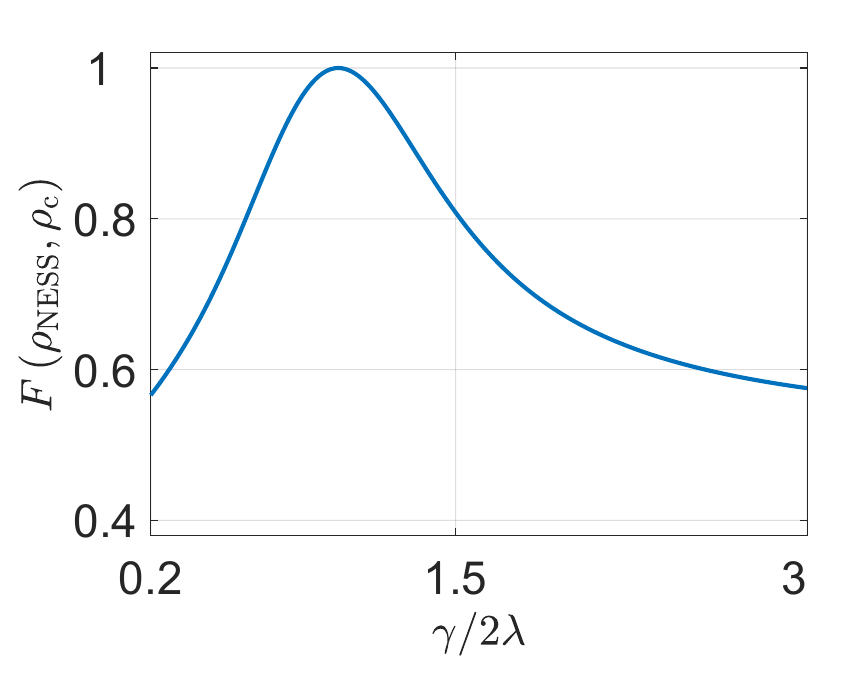}
\caption{Plot of $F\left( \protect\rho _{\mathrm{NESS}},\protect\rho _{\mathrm{c}}\right) $ as a function of $\protect\gamma $. The strength of the
external field is fixed at $\protect\lambda =0.5$. The configuration of the
system is shown in Fig. \protect\ref{fig_illu}. Notably, a peak is observed
at $\protect\gamma =2\protect\lambda $, which corresponds to the EP of the
effective non-Hermitian Hamiltonian $\mathcal{H}_{\mathrm{spin}}$. It is
worth mentioning that slight deviations from $1$ do not significantly impact
the NESS. This observation indicates the existence of a parameter window
that allows for magnetization induced by local dissipation.} \label{fig5}
\end{figure}

Now let us further investigate whether this conclusion holds when the system
parameters are not finely tuned. First, we consider the case where $\gamma $
deviates from $\gamma _{c}=\lambda /2$. We plot Fig. \ref{fig5}, which shows
$F\left( \rho \left( t\rightarrow \infty \right) ,\rho _{\mathrm{c}}\right) $
as a function of $\gamma $. It can be observed that the final steady state
is almost unaffected as $\gamma $ deviates slightly from $\gamma _{c}$.
However, when $\gamma \ll \gamma _{c}$, there are no EP and complex energy
in $\mathcal{H}_{\mathrm{spin}}$. In this case, the state initialized in the
subspace $\left\{ \left\vert G_{n}\right\rangle \right\} $ will not tend to
a definite state but instead oscillates between different eigenenergies,
resulting in a periodic oscillation of the physical observables. This can be
seen from the Fig. \ref{fig4}(a). Consequently, the density matrix driven by
$\mathcal{H}_{\mathrm{spin}}$ will exhibit distinct dynamics from the
quantum jump operator which forces the spin along the $y$-direction.
Combining both effects, the NESS deviates from $\rho _{\mathrm{c}}$. On the
other hand, when $\gamma \gg \gamma _{c}$, $\mathcal{H}_{\mathrm{spin}}$
drives all the spins to the down states since the eigenstate $\left\vert
\Downarrow \right\rangle $ has the largest imaginary part. However, this
contradicts the action of the quantum jump operator. As a consequence of the
parameter deviation, the final state is no longer a product state with a
definite direction but a mixed state that loses some of its coherence.
\begin{figure}[tbh]
\centering\includegraphics[width=0.45\textwidth]{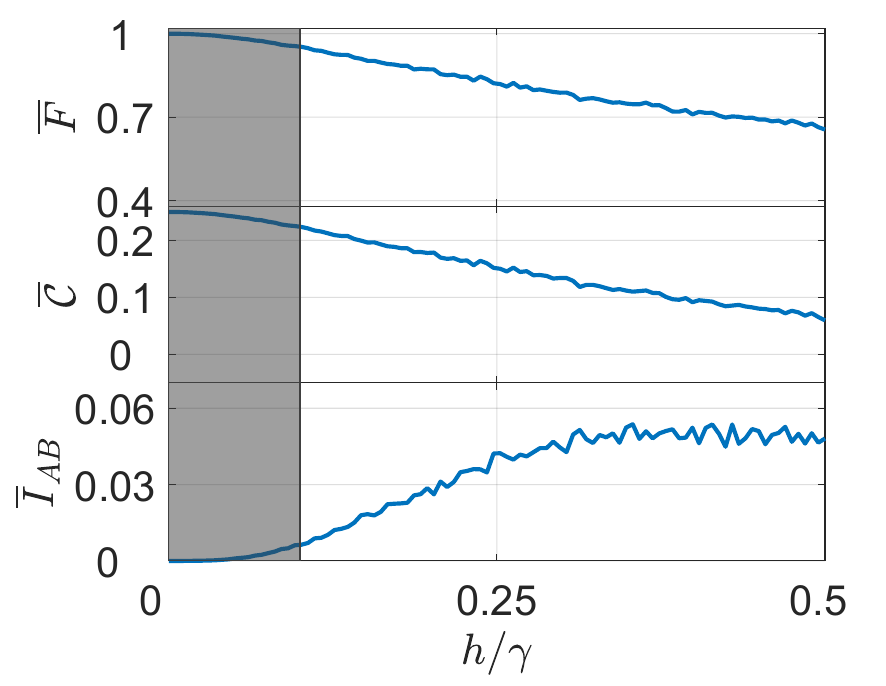}
\caption{Numerical simulations of $\overline{F}\left( \protect\rho _{\mathrm{NESS}},\protect\rho _{\mathrm{c}}\right)$, $\overline{\mathcal{C}}\left( t\rightarrow
\infty \right)$, and $\overline{I}_{AB}\left( t\rightarrow \infty \right)$ as functions of the disorder strength $h$. The system parameters are fixed at $\protect\lambda /\protect\gamma =0.5$, and $J_{ij}/\protect\gamma =1$. The structure of the system is
depicted in Fig. \protect\ref{fig_illu}. The average Uhlmann fidelity $\overline{F}\left( \protect\rho _{\mathrm{NESS}},\protect\rho _{\mathrm{c}}\right) $, average correlator $\overline{\mathcal{C}}\left( t\rightarrow
\infty \right) $, and average quantum mutual information $\overline{I}_{AB}\left( t\rightarrow \infty \right) $ are calculated by averaging over
1000 disorder configurations. The simulations demonstrated that the
realization of $\protect\rho _{\mathrm{c}}$ remains unaffected by the
specific system configuration and is immune to weak disorder, as indicated
by the gray shaded region. This property is advantageous for observing the
magnetization induced by single local dissipation in experimental setups.} %
\label{fig6}
\end{figure}

Besides the deviation from the EP, another factor influencing the
success of the scheme is the presence of disorder. In the experiment, our
proposal can be realized in a cold atom system, particularly in the Rydberg
atom quantum simulator. \cite{Smith2016,Zhang2017,Marcuzzi2017,Shibata2020,Mondragon-Shem2021} The system can be subjected to disorder
through external fields, such as electric or magnetic fields. Fluctuations
or variations in the strength and direction of these fields can impact the
energy levels and dynamics. It is crucial to examine the system's robustness
to disorder. To achieve this, we introduce disorder by considering a random
magnetic field in the $z$ direction. The modified system Hamiltonian is
given as
\begin{equation}
H_{\text{\textrm{spin}}}^{\mathrm{d}}=H_{\text{\textrm{spin}}}+H^{\mathrm{d}%
},
\end{equation}%
with
\begin{equation}
H^{\mathrm{d}}=\sum_{i}h_{i}s_{i}^{z},
\end{equation}%
where $h_{i}$ represents a random number within the range $(-h,$ $h)$.
Clearly, $H^{\mathrm{d}}$ breaks the SU(2) symmetry of $H_{\text{\textrm{spin%
}}}^{\mathrm{d}}$ and hence prevents the formation of the subpace of $%
\left\{ \left\vert G_{n}\right\rangle \right\} $. Although exact SU(2)
symmetry is spoiled, it can be inferred that the directed evolution in the $%
\left\{ \left\vert G_{n}\right\rangle \right\} $ subspace may still exist
under weak disorder. In Fig. \ref{fig6}, we perform the numerical simulation
to examine the average Uhlmann fidelity $\overline{F}\left( \rho _{\mathrm{%
NESS}},\rho _{\mathrm{c}}\right) $, average correlator $\overline{\mathcal{C}%
}\left( t\rightarrow \infty \right) $ and average quantum mutual information
$\overline{I}_{AB}\left( t\rightarrow \infty \right) $. The results
demonstrate that a small distribution of $h$ does not induce a transition in
the final state as the degenerate subspace $\left\{ \left\vert
G_{n}\right\rangle \right\} $ is approximately preserved, as manifested by
the behavior of $\overline{\mathcal{C}}\left( t\rightarrow \infty \right) $
and $\overline{I}_{AB}\left( t\rightarrow \infty \right) $ in the grey
shaded region. However, when $h$ is large enough to completely destroy the
SU(2) symmetry, the dynamics of EP cannot be maintained as the degenerate
subspace ceases to exist. In such a scenario, the action of $\mathcal{H}_{%
\mathrm{spin}}$ and the quantum jump operator $\widetilde{L}$ exhibit
distinct dynamics, leading to the collapse of the final ferromagnetic state.

\section{Summary}

\label{summary} In conclusion, we have demonstrated that the critical
non-Hermitian system accurately captures the long-term dynamics of the open
quantum system. Specifically, the master equation of the open quantum system
can be rephrased as a stochastic average over individual trajectories, which
can be numerically evolved as pure states over time. Each trajectory's
evolution is determined by the SSE. There are two types of probabilistic
evolution: a non-unitary evolution driven by the effective non-Hermitian
Hamiltonian and a state projection determined by the quantum jump operator.
The trade-off between these two evolutions determines the final NESS. For
the non-Hermitian Hamiltonian, a definite final evolved state can be
achieved if the system possesses the EP or an imaginary energy level. In the
former case, the evolved state is forced towards the coalescent state, while
in the latter case, it approaches the eigenstate with the maximum value of
the imaginary energy level. If the final evolved state coincides with the
state under the quantum jump operation, then the NESS of the open quantum
system is identical to the coalescent state of the effective non-Hermitian
Hamiltonian. Furthermore, we apply this mechanism to the open quantum spin
system and find that local critical dissipation can induce a high-order EP
in the effective non-Hermitian ferromagnetic Heisenberg system. The
dimension of the degenerate subspace determines the order of the EP. The
corresponding coalescent state represents all the spins aligned in parallel
to the $y$-direction. From a dynamical perspective, when the initial state
is prepared within the degenerate subspace, the EP dynamics force all the
spins to align in the y-direction regardless of the initial spin
configuration. On the other hand, the quantum jump operator rotates the spin
that passes the first site to align with the direction of the coalescent
state. Both actions of the two probabilistic propagations are identical,
leading to the NESS being the coalescent state. The realization of this type
of NESS is immune to weak disorder and holds within a certain range of
system parameters. These findings serve as the building blocks for
understanding critical open quantum systems from both theoretical and
experimental perspectives.

\acknowledgments X.Z.Z. gratefully acknowledge Y.L. for hosting and
generously providing the necessary resources for this study. We acknowledge
the support of the National Natural Science Foundation of China (Grants No.
12275193, No. 12225507, and No. 12088101), and NSAF (Grant No. U1930403).

\section{Appendix}

\subsection{EP dynamics of two-level system}

\label{EP_two} In this subsection, we analyze the EP dynamics in a
non-Hermitian two-level system. The Hamiltonian, given by%
\begin{equation}
\mathcal{H}=\lambda s^{x}-\frac{i\gamma }{2}s^{+}s^{-},
\end{equation}%
is non-Hermitian due to the dissipation channel. In the basis of $%
\{|\uparrow \rangle ,$ $|\downarrow \rangle \}$, the matrix form of $%
\mathcal{H}$ is expressed as%
\begin{equation}
\mathcal{H}=\frac{1}{2}\left(
\begin{array}{cc}
-i\gamma /2 & \lambda \\
\lambda & i\gamma /2%
\end{array}%
\right) -\frac{i\gamma }{4}I,
\end{equation}%
where $\frac{i\gamma }{4}I$ is a constant term that does not affect the
relative probability of populating the two different energy states. The two
eigenstates of $\mathcal{H}$ coalesce at the EP when $\lambda =\gamma /2$.
The corresponding coalescent state is $|\psi _{\mathrm{c}}\rangle
=|y,+\rangle =\frac{1}{\sqrt{2}}\left( 1,\text{ }i\right) ^{T}$ which also
represents the eigenstate of $s^{y}$ with eigenenergy $1/2$. Now, let us
turn focus on the system propagator $\mathcal{U}_{2}$. Due to the nilpotent
matrix property of $\mathcal{H}$, i.e., $\mathcal{H}^{2}=0$, $\mathcal{U}%
_{2} $ simplifies to
\begin{equation}
\mathcal{U}_{2}=e^{-i\mathcal{H}t}=e^{-\gamma t/4}\left[ 1-\frac{i\lambda t}{%
2}\left(
\begin{array}{cc}
-i & 1 \\
1 & i%
\end{array}%
\right) \right] .
\end{equation}%
For an arbitrary initial state $|\psi \left( 0\right) \rangle =\left( a,%
\text{ }b\right) ^{T}$, the evolved state can be given as
\begin{equation}
|\psi \left( t\right) \rangle =e^{-\gamma t/4}\left(
\begin{array}{c}
a-i\lambda t/2 \\
b+\lambda t/2%
\end{array}%
\right) ,
\end{equation}%
As time tends to infinity, $|\psi \left( t\right) \rangle $ normalized in
terms of Dirac probability approaches $|\psi \left( \infty \right) \rangle
=|\psi _{\mathrm{c}}\rangle $.

\subsection{NESS of an open two-level system}

\label{NESS_two} In this subsection, we derive the NESS of the two-level
open system under consideration. The LME describing the open system dynamics
is given by%
\begin{equation}
\frac{\text{\textrm{d}}\rho }{\text{\textrm{d}}t}=-i\lambda \lbrack
s^{x},\rho ]-\frac{\gamma }{2}\left( s^{+}s^{-}\rho +\rho s^{+}s^{-}\right)
+\gamma s^{-}\rho s^{+}.
\end{equation}%
By applying the Choi-Jamio\l kowski isomorphism, the LME can be written as
an equivalent form:%
\begin{equation}
\frac{\text{\textrm{d}}\rho }{\text{\textrm{d}}t}\equiv \widetilde{\mathcal{L%
}}\rho ,
\end{equation}%
where the vectorized density matrix $|\rho \rangle =\sum_{m,n}\rho
_{m,n}|m\rangle \otimes |n\rangle $ represents the density matrix in the
double space. The Liouvillian superoperator is given by%
\begin{eqnarray}
\widetilde{\mathcal{L}} &=&-i\lambda \left( s^{x}\otimes I-I\otimes
s^{x}\right) +\gamma s^{-}\otimes s^{-}  \notag \\
&&-\frac{\gamma }{2}\left( s^{+}s^{-}\otimes I+I\otimes s^{+}s^{-}\right) .
\end{eqnarray}%
The matrix representation of $\widetilde{\mathcal{L}}$ can be expressed as%
\begin{equation}
\widetilde{\mathcal{L}}=\left(
\begin{array}{cccc}
-\gamma & i\frac{\lambda }{2} & -i\frac{\lambda }{2} & 0 \\
i\frac{\lambda }{2} & -\frac{1}{2}\gamma & 0 & -i\frac{\lambda }{2} \\
-i\frac{\lambda }{2} & 0 & -\frac{1}{2}\gamma & i\frac{\lambda }{2} \\
\gamma & -i\frac{\lambda }{2} & i\frac{\lambda }{2} & 0%
\end{array}%
\right) .
\end{equation}%
The complete spectrum of the Liouvillian superoperator $\mathcal{L}$ can be
obtained by solving the eigen-equation: $\widetilde{\mathcal{L}}|\rho
_{k}\rangle =\varepsilon _{k}|\rho _{k}\rangle $, where $k$ represents the
eigenvalue and $|\rho _{k}\rangle $ denotes its corresponding eigenmatrix.
The NESS is unique and corresponds to the eigenvalue $\varepsilon _{k}=0$.
Straightforward algebra reveals that the corresponding eigenmatrix is given
by%
\begin{equation}
\rho _{\mathrm{NESS}}=\left(
\begin{array}{cc}
\frac{\lambda ^{2}}{2\lambda ^{2}+\gamma ^{2}} & -i\frac{\lambda \gamma }{%
2\lambda ^{2}+\gamma ^{2}} \\
i\frac{\lambda \gamma }{2\lambda ^{2}+\gamma ^{2}} & \frac{\lambda
^{2}+\gamma ^{2}}{2\lambda ^{2}+\gamma ^{2}}%
\end{array}%
\right) .
\end{equation}

\subsection{Non-Hermitian Heisenberg model and EP dynamics}

\label{NHH}

\subsubsection{model and EP}

\label{ME} In this subsection, we analyze the non-Hermitian Heisenberg model
with a local dissipation channel and identify the EP. According to the main
text, the Hamiltonian of the effective non-Hermitian Heisenberg model in the
LME under an external field is given by%
\begin{equation}
\mathcal{H}_{\mathrm{spin}}=H_{\mathrm{spin}}+H_{\mathrm{ec}},
\end{equation}%
where
\begin{equation}
H_{\mathrm{spin}}=-\frac{1}{2}\sum_{i,j\neq
i}J_{ij}(s_{i}^{+}s_{j}^{-}+s_{i}^{-}s_{j}^{+}+2s_{i}^{z}s_{j}^{z}),
\end{equation}%
and
\begin{equation}
H_{\mathrm{ec}}=\sum_{\left\{ i\right\} }\lambda _{i}\mathbf{h}\cdot \mathbf{%
s}_{i}-\frac{i}{2}\sum_{i}\Gamma _{i}s_{i}^{+}s_{i}^{-}.
\end{equation}%
$H_{\mathrm{ec}}$ can be deemed as the external complex magnetic filed.
Here, $\left\{ i\right\} $ represents a set of multiple local sites that are
subjected to the local complex fields. The presence of inhomogeneous
magnetic fields breaks the $SU(2)$ symmetry, i.e., $\left[ s^{\pm },\mathcal{%
H}_{\mathrm{spin}}\right] \neq 0$. However, $H_{\mathrm{spin}}$ and $H_{%
\mathrm{ec}}$, commute with each other when the homogeneous magnetic field
and dissipation are applied, i.e., $\lambda _{i}=\lambda $ and $\Gamma
_{i}=\gamma $. Although these Hamiltonians share common eigenstates, the
properties of the ground states are unclear due to the non-Hermitian nature
of $H_{\mathrm{ec}}$. This poses a challenge to perturbation theory in
Hermitian quantum mechanics since the omission of high-order corrections
cannot be guaranteed in the complex regime. To simplify the analysis%
, we consider $\lambda _{i}=\lambda \delta _{i,1}$, and $\Gamma _{i}=\gamma
\delta _{i,1}$ from this point onwards. To proceed, we introduce a
similarity transformation $\mathcal{S}_{1}=\prod\nolimits_{j}\mathcal{S}%
_{1}^{j}$, where $\mathcal{S}_{1}^{j}=e^{-i\theta s_{j}^{y}}$ represents a
counter-clockwise spin rotation in the $s_{x}$-$s_{z}$ plane around the $%
s_{y}$-axis by an angle $\theta $. Here $\theta $ is a complex number
dependent on the strength of the complex field, given by $\theta =\tan
^{-1}\left( 2\lambda /i\gamma \right) $. It is important to note that the
spin rotation $\mathcal{S}_{1}^{j}$ is valid at arbitrary $\gamma $ except
at EP of $H_{\mathrm{ec}}$ ($\lambda =\gamma /2$), where $H_{\mathrm{ec}}$
takes a non-diagonalizable Jordan block form. Under the spin-rotation, the
transformed Hamiltonian is as follows:
\begin{eqnarray}
\overline{\mathcal{H}}_{\mathrm{spin}} &=&\overline{H}_{\mathrm{spin}}+%
\overline{H}_{\mathrm{ec}}, \\
\overline{H}_{\mathrm{spin}} &=&-\frac{1}{2}\sum_{i,j\neq i}J_{ij}(\tau
_{i}^{+}\tau _{j}^{-}+\tau _{i}^{-}\tau _{j}^{+}+2\tau _{i}^{z}\tau
_{j}^{z}), \\
\overline{H}_{\mathrm{ec}} &=&\sqrt{\lambda ^{2}-\gamma ^{2}/4}\tau _{1}^{z}-%
\frac{i\gamma }{4},
\end{eqnarray}%
where the new set of operators $\tau _{j}^{\pm }=(\mathcal{S}%
_{1}^{j})^{-1}s_{j}^{\pm }\mathcal{S}_{1}^{j}$ and $\tau _{j}^{z}=(\mathcal{S%
}_{1}^{j})^{-1}s_{j}^{z}\mathcal{S}_{1}^{j}$ also satisfies the Lie algebra.
We omit the overall decay factor $-i\gamma /4$ which served as the energy
base has no effects on the subsequent evolution. Specifically, they obey the
commutation relations $[\tau _{i}^{z},\tau _{j}^{\pm }]=\pm \tau _{i}^{\pm
}\delta _{ij}$ and $[\tau _{i}^{+},\tau _{j}^{-}]=2\tau _{i}^{z}\delta _{ij}$%
. It is important to note that $\tau _{j}^{\pm }\neq (\tau _{j}^{\mp
})^{\dagger }$ due to the complex rotation angle $\theta $. We consider the
eigenstates of the operator $\sum_{i}s_{i}^{z}$, denoted as $\left\{
\left\vert \psi _{n}\right\rangle \right\} $, which represent possible spin
configurations along the $+z$ direction. Under the biorthogonal basis of $\{%
\mathcal{S}^{-1}\left\vert \psi _{n}\right\rangle \}$ and $\{\mathcal{S}%
^{\dagger }\left\vert \psi _{n}\right\rangle \}$, the matrix form of $%
\overline{\mathcal{H}}_{\mathrm{spin}}$ is Hermitian for $\lambda >\gamma /2$
except a complex energy base. Although the presence of the local complex
field breaks the SU(2) symmetry of the system, as indicated by $[\overline{H}%
_{\mathrm{ec}},$ $\overline{\mathcal{H}}_{\mathrm{spin}}]\neq 0$, the
entirely real spectrum remains without symmetry protection. When $%
\lambda =\gamma /2$, the transformation of $\mathcal{S}_{1}$ is ill-defined, indicating that $H_{\mathrm{ec}}$ is non-diagonalizable, which corresponds to the presence of an EP. In principle, the EP of $\mathcal{H}_{\mathrm{spin}}$ or $\overline{%
\mathcal{H}}_{\mathrm{spin}}$ may not coincide with the EP of $H_{\mathrm{ec}%
}$ at $\lambda =\gamma /2$. In the following, we will demonstrate that $%
\mathcal{H}_{\mathrm{spin}}$ and $H_{\mathrm{ec}}$ exhibit the same EP
behavior within the framework of perturbation theory.

The Hermiticity of the matrix representation of $\overline{\mathcal{H}}_{%
\mathrm{spin}}$ allows us to apply various approximation methods in quantum
mechanics. As$\ \gamma $ approaches $2\lambda $, the value of $\sqrt{\lambda
^{2}-\gamma ^{2}/4}$ becomes small, allowing $\overline{H}_{\mathrm{ec}}$ in
the new frame to be treated as a weak perturbation. Our focus is on the
influence of $\overline{H}_{\mathrm{ec}}$ on the ground states $\left\{
\left\vert G_{n}^{\prime }\right\rangle \right\} $ of $\overline{H}_{\mathrm{%
spin}}$. Due to the properties of the ferromagnetic spin system $\overline{H}%
_{\mathrm{spin}}$, the ground states $\left\{ \left\vert G_{n}^{\prime
}\right\rangle \right\} $ exhibit $(N+1)$ fold-degeneracy and can be
expressed as
\begin{equation}
\left\vert G_{n}^{\prime }\right\rangle =(\sum_{i}\tau
_{i}^{-})^{n-1}\left\vert \Uparrow \right\rangle ^{\prime }\text{ }\left(
n=1,\text{ }2\text{ }...\text{ }N+1\right) \text{, }
\end{equation}%
where
\begin{equation}
\left\vert \Uparrow \right\rangle ^{\prime }=(\mathcal{S}_{1})^{-1}\left%
\vert \Uparrow \right\rangle \text{, and }\left\vert \Uparrow \right\rangle
=\prod\limits_{i=1}^{N}\left\vert \uparrow \right\rangle _{i}\text{.}
\end{equation}%
$\left\vert G_{n}^{\prime }\right\rangle $ is also the eigenstate of $%
\mathbf{\tau }^{2}=\sum_{i}\mathbf{\tau }_{i}^{2}$ with $\tau =N/2$, where $%
N $ denotes the total number of spins. Noticeably, the presence of
degenerate ground states is irrelevant to the system's structure. This
property can be observed in other types of systems as well \cite%
{Heisenberg1928,Yang1966}. Following the principles of degenerate
perturbation theory, the eigenvalues up to first order can be determined by
the matrix representation of $\overline{H}_{\mathrm{ec}}$ within the
subspace spanned by $\left\{ \left\vert G_{n}^{\prime }\right\rangle
\right\} $. For simplicity, we refer to the corresponding perturbed matrix
as $W^{\prime }$, with elements given by $W_{m,n}^{\prime }=\langle
\overline{G}_{m}^{\prime }|\overline{H}_{\mathrm{ec}}|G_{n}^{\prime }\rangle
$. The biorthogonal left eigenvectors are denoted as $\{\langle \overline{G}%
_{m}^{\prime }|\}$ and can be expressed as
\begin{equation}
\langle \overline{G}_{m}^{\prime }|=\left\langle \Uparrow \right\vert
\mathcal{S}_{1}(\sum_{i}\tau _{i}^{+})^{m-1}\text{ }\left( m=1,\text{ }2%
\text{ }...\text{ }N+1\right) .
\end{equation}%
Two important points are highlighted: (i) Owing to the Hermiticity of the
matrix $W^{\prime }$, higher-order corrections can be safely disregarded as$%
\ \gamma $ approaches $2\lambda $. (ii) When a homogeneous magnetic field is
applied, $\left[ \overline{H}_{\mathrm{ec}},\text{ }\overline{\mathcal{H}}_{%
\mathrm{spin}}\right] =0$, enabling the decomposition of $\overline{\mathcal{%
H}}_{\mathrm{spin}}$ into block matrices based on the eigenvectors of $%
\mathbf{\tau }^{2}$. Consequently, the eigenvalues of $W^{\prime }$ comprise
the energies of the ground state and $N$ excited states of $\overline{%
\mathcal{H}}_{\mathrm{spin}}$. After straightforward algebras, the entry of
the matrix can be obtained as
\begin{equation}
\ W_{m,n}^{\prime }=\sqrt{\lambda ^{2}-\gamma ^{2}/4}[\left( N/2-m+1\right)
\delta _{m,n}]/N,
\end{equation}%
where the factor $1/N$ arises from the translation symmetry of the ground
state $\{\left\vert G_{n}\right\rangle ^{\prime }\}$. By performing the
transformation $W=U\mathcal{S}_{1}W^{\prime }(\mathcal{S}_{1})^{-1}U^{-1}$ ($%
W_{m,n}=\langle \widetilde{G}_{m}|UH_{\mathrm{ec}}U^{-1}|\widetilde{G}%
_{n}\rangle $ with $|\widetilde{G}_{n}\rangle =U\left\vert
G_{n}\right\rangle $), the matrix element of $W$ can be expressed as
\begin{eqnarray}
W_{m,n} &=&\sqrt{\left( N+1-m\right) m}[\left( \lambda -\gamma /2\right)
\delta _{m+1,n}  \notag \\
&&+\left( \lambda +\gamma /2\right) \delta _{m,n+1}]/2N.
\end{eqnarray}%
When $\lambda =\gamma /2$, it reduces to a Jordan block form, and an EP of
order $N+1$ occurs. The corresponding coalsecent is $|\psi _{\mathrm{c}%
}\rangle =\prod\nolimits_{j}e^{-i\frac{\pi }{2}s_{j}^{x}}\left\vert
\Downarrow \right\rangle $. It is worth mentioning that if we express $H_{%
\mathrm{ec}}$ in the basis of $\left\{ \left\vert G_{n}\right\rangle
\right\} $, it describes a $\mathcal{PT}$-symmetric hypercube graph of $N+1$
dimension \cite{Zhang2012}. The EP also emerges when $\lambda =\gamma /2$.

\subsubsection{high order EP dynamics}

\label{HEP} In this subsection, our objective is to generate a saturated
ferromagnetic state where all local spins (or conduction electron spins) are
aligned parallel to the $y$-direction. The non-Hermitian Heisenberg
Hamiltonian is represented by Eq. (\ref{non_spin}) in the main text.
Considering the EP $\lambda =\gamma /2$ within the subspace $\{|\widetilde{G}%
_{n}\rangle \}$, the matrix form of $W$ can be expressed as
\begin{equation}
W_{m,n}=\lambda \sqrt{\left( N+1-m\right) m}\delta _{m,n+1}/N,
\end{equation}
which corresponds to a Jordan block of dimension $N+1$. The coalescent
eigenstate is $|\widetilde{G}_{N+1}\rangle $. It is important to note that $%
W $ is a nilpotent matrix with order $\left( N+1\right) $ meaning that $%
\left( W\right) ^{N+1}=0$. The element of matrix $W^{k}$ can be given as
\begin{equation}
\left( W^{k}\right) _{mn}=[\prod\limits_{p=m+1-k}^{m}p\left( N+1-p\right)
]^{1/2}(\frac{\lambda }{N})^{k}\delta _{m,n+k},  \label{ele_W}
\end{equation}%
where $k<m+1$. Our attention now shifts to the dynamics of the critical
matrix $W $, and the evolution of states within this subspace is governed by
the propagator $\mathcal{U=}e^{-iWt}$. Utilizing Eq. (\ref{ele_W}), we can
derive the elements of the propagator $\mathcal{U}$ as follows:
\begin{eqnarray}
\mathcal{U}_{m,n} &=&\delta _{mn}+\left( \frac{-it\lambda }{N}\right) ^{m-n}%
\frac{h\left( m-n\right) }{\left( m-n\right) !}  \notag \\
&&\times \lbrack \prod\limits_{p=n+1}^{m}p\left( N+1-p\right) ]^{1/2},
\end{eqnarray}%
where $h\left( x\right) $ is a step function defined as $h\left( x\right) =1$
$\left( x>0\right) ,$ and $h\left( x\right) =0$ $\left( x<0\right) $.
Considering an arbitrary initial state $\sum_{n}c_{n}\left( 0\right) |%
\widetilde{G}_{n}\rangle $, the coefficient $c_{m}\left( t\right) $ of the
evolved state is given by
\begin{eqnarray}
c_{m}\left( t\right) &=&c_{m}\left( 0\right) +\sum_{n\neq m}\left( \frac{%
-it\lambda }{N}\right) ^{m-n}\frac{h\left( m-n\right) }{\left( m-n\right) !}
\notag \\
&&\times \lbrack \prod\limits_{p=n+1}^{m}p\left( N+1-p\right)
]^{1/2}c_{n}\left( 0\right) .  \label{evolve_c}
\end{eqnarray}%
It is evident that regardless of the initial state chosen, the coefficient $%
c_{N+1}\left( t\right) $ of evolved state always contains the highest power
of time $t$. As time progresses, the component $c_{N+1}\left( t\right) $ of
the evolved state overwhelms the other components, ensuring the final state
is coalescent state $|\psi _{\mathrm{c}}\rangle =\prod\nolimits_{j}e^{-i%
\frac{\pi }{2}s_{j}^{x}}\left\vert \Downarrow \right\rangle $ under the
Dirac normalization. The different types of initial states only determine
how the total probability of the evolved state increases over time and the
relaxation time for it to evolve towards the coalescent state.

\subsection{NESS of the open quantum spin system subjected to a local
magnetic field}

\label{NESS_local} In this subsection, we demonstrate that the critical
density matrix $\rho _{\mathrm{c}}=|\psi _{\mathrm{c}}\rangle \langle \psi _{%
\mathrm{c}}|$ is also the NESS of the open quantum spin system. The dynamics
of the open quantum spin system under consideration is governed by LME,
expressed as:
\begin{eqnarray}
\frac{\text{\textrm{d}}\rho }{\text{\textrm{d}}t} &=&-i(\mathcal{H}_{\mathrm{%
spin}}\rho -\rho \mathcal{H}_{\mathrm{spin}}^{\dagger })  \notag \\
&&+\gamma \left( s_{1}^{x}-is_{1}^{z}\right) U\rho U^{-1}\left(
s_{1}^{x}+is_{1}^{z}\right)  \notag \\
&\equiv &\mathcal{L}\rho ,  \label{spin_l}
\end{eqnarray}%
where $U$ is defined as the product of operators $U=\prod\nolimits_{j}e^{-i%
\pi s_{j}^{x}/2}$ and%
\begin{eqnarray}
\mathcal{H}_{\mathrm{spin}} &=&H_{\mathrm{spin}}+H_{\mathrm{ec}}, \\
H_{\mathrm{spin}} &=&-\frac{1}{2}\sum_{i,j\neq
i}J_{ij}(s_{i}^{+}s_{j}^{-}+s_{i}^{-}s_{j}^{+}+2s_{i}^{z}s_{j}^{z}), \\
H_{\mathrm{ec}} &=&\lambda s_{1}^{x}-\frac{i\gamma }{2}s_{1}^{+}s_{1}^{-},
\end{eqnarray}%
Here $\lambda =\gamma /2$ is assumed when $\mathcal{H}_{\mathrm{spin}}$ is
at EP. Next, we substitute $\rho _{\mathrm{c}}=|\psi _{\mathrm{c}}\rangle
\langle \psi _{\mathrm{c}}|$ into the above equation. Recalling that the $%
|\psi _{\mathrm{c}}\rangle =\prod\nolimits_{j}e^{-i\frac{\pi }{2}%
s_{j}^{x}}\left\vert \Downarrow \right\rangle $, we can readily deduce that $%
\mathcal{H}_{\mathrm{spin}}|\psi _{\mathrm{c}}\rangle =-i\gamma /4|\psi _{%
\mathrm{c}}\rangle $, resulting in
\begin{equation}
-i(\mathcal{H}_{\mathrm{spin}}\rho _{\mathrm{c}}-\rho _{\mathrm{c}}\mathcal{H%
}_{\mathrm{spin}}^{\dagger })=-\frac{\gamma }{2}\rho _{\mathrm{c}}.
\end{equation}%
Applying $\left( s_{1}^{x}-is_{1}^{z}\right) U$ to $|\psi _{\mathrm{c}%
}\rangle $ yields $\frac{1}{\sqrt{2}}|\psi _{\mathrm{c}}\rangle $. Thus, we
can conclude that $\mathcal{L}\rho _{\mathrm{c}}=0$, demonstrating that $%
\rho _{\mathrm{c}}$ is indeed the NESS $\rho _{\mathrm{NESS}}$ of the open
quantum spin system.

\end{document}